\documentclass[preprint,preprintnumbers,amsmath,amssymb]{revtex4}
\usepackage{graphicx} 
\usepackage{setspace} 
\usepackage{dcolumn} 
\usepackage{bm} 
\usepackage{natbib}
\usepackage{hyperref}
\usepackage{tabularx}
\hypersetup{colorlinks=true,linkcolor=blue,citecolor=blue,urlcolor=blue}
\usepackage{booktabs}

\begin{document}

\title{Quantum Materials Group Annual Report 2022}
\author{P. Kumari, S. Rani, S. Kar, T. Mukherjee, S. Majumder, K. Kumari, S. J. Ray}
\email{ray@iitp.ac.in, ray.sjr@gmail.com}
\affiliation{Department of Physics, Indian Institute of Technology Patna, Bihta 801106, India}

\begin{abstract}
The Quantum Materials group at Indian Institute of Technology Patna is working on a range of topics relating to nanoelectronics, spintronics, clean energy and memory design etc. The PI has past experiences of working extensively with superconducting systems like cuprates \cite{prl1, prb4}, ruthanate \cite{prb2}, pnictide \cite{pss1, prb1}, thin film heterostructures \cite{prb3, pc1} etc and magnetic recording media \cite{apl1, jap1} etc. In this report, we have summarised the ongoing works in our group. We explored a range of functional materials like two-dimensional materials, oxides. topological insulators, organic materials etc. using a combination of experimnetal and computational tools. Some of the useful highlights are as follows: (a) tuning and control of the magnetic and electronic state of 2D magentic materials with rapid enhancement in the Curie temperature, (b) Design and detection of single electron transistor based nanosensors for the detection of biological species with single molecular resolution, (c) Observation of non-volatile memory behaviour in the hybrid structures made of perovskite materials and 2D hybrids. The results offer useful insight in the design of nanoelectronic architecrures for diverse applications.
\end{abstract}


\maketitle

\section{Nanosensing}
The discovery of 2D materials has opened up various areas for nanoelectronic application. 
A single electron transistor (SET) is a source-quantum dot-drain structure where the quantised conduction through the dot can be controlled by a capacitively coupled gate. The quantised energy levels in the quantum dot (QD) give rise to well defined conductance plateaus, with their positions changing at specific source-drain bias leading to a coulomb blockade state. 

Due to the larger surface to volume ratio, 2D materials attracted significant attention for sensing applications, which is reflected in its drastic change of resistance on the adsorption of an alien molecule. In the close proximity of an external molecule, the charge carrier concentration on a 2D material changes significantly. This property has been exploited to use 2D layered materials like graphene \cite{cej1}, borophene \cite{pccp6} and MoS2 for sensing applications. Depending upon the sign of the charge transfer between the molecule and the 2D material, the molecule near the layered material can work as a donor or acceptor. Owing to the 2D nature of such materials, the effect of surface dopants is extremely perceptive which has been demonstrated in achieving extreme detection efficiency upto 1 ppb. 

In our recent work, we have developed various single electron transistor (SET) devices with 2D materials as island and studied their usefulness as nanosensor. DNA and RNA detection were studied for the case of graphene and hexagonal boron nitride based SET \cite{carbon1, icee1}. It was observed that addition of a secondary gate electrode offers better control of the detection efficiency \cite{jap8}. Earlier, we have designed various SET devices for chemical detection using a variety of layers like graphene, MoS$_2$ \cite{sab1}, phosphorene \cite{jpcm1}, C$_3$N etc \cite{carbon2} and in various engineered nanostructures \cite{itn1, ne1, jap7, jap6, jap5, jap4, jap3, jap2}. The biosensing behaviour of a C$_3$N nanoribbon can be found here \cite{pccp1, aip2} and Proximity-induced colossal conductivity modulation was observed in phosphorene towards the detection of organic molecules \cite{pra1}. Silicon based single electron devices have been used for electron pumping operation, that is useful for quantum metrology \cite{pra2, cpem1, wolte1, cpem2, prx1}.

\section{Spintronics}

The discovery of atomically thin two dimensional materials has brought a paradigm shift in the development of 2D nanoelectronic and spintronic devices. Today, a wide variety of such materials ranging from graphene, which is a semi-metal, to semiconductors and insulators have emerged with unique electronic properties. Specifically, graphene and other 2D crystals have already shown their utility in spintronic applications \cite{apr1, jpcc1, aip3, mtp1, aps4}. A recent entry into this area has been the class of 2D magnets, which defied the predictions of Mermin-Wagner theorem, which states that finite temperature thermal excitations are enough to destroy magnetic order in two dimension. Despite this, the feasibility of 2D magnets was shown by first-principles calculations in several 2D materials, which can be considered as ground states at absolute zero temperature. Several recent experiments showed the existence of 2D magnetism in several predicted materials such as CrI$_3$, VSe$_2$ and Cr$_6$Ge$_2$Te$_6$. This validity has encouraged exploring the feasibility of magnetic phases in more 2D crystals. However, the majority of the predicted materials exhibit ordering temperatures much below room temperature and a key challenge remains today is to find ways to increase the ordering temperature above room temperature and engineer phases locally. In particular, such control is necessary for device realisation and applicability of the magnetic phases in spintronic and magneto-electronic switches. The conventional strategy involves chemical doping and defect generation in 2D crystals for creating local magnetic moments \cite{pccp3, pccp2, aps13}. However, to dynamically establish and control magnetism in materials, it is essential to use physical stimuli such as electric field or strain that can be locally applied at a nanoscopic device scale \cite{pccp5}. 

In this direction, we have recently uncovered the family of 2D transition metal oxychlorides with the general formula MOCl (M = Ti, V, Fe, Cr). This type of magnetic material, while being a semiconductor, is promising for integration into electronic circuits. However, like other 2D semiconductor magnets, ferromagnetism in CrOCl only exists at low temperatures, much below 200 K. This makes it highly interesting to devise new ways of enhancing the interatomic exchange interaction for designing and tuning magnetic phases. While electric field control of the carrier concentration in graphene and other 2D crystals is one way to tune the interatomic exchange, the exceptional resilience of such crystals to strain also provides novel ways to engineer desired electronic structures and enable flexible 2D spintronic devices. For CrOCl, we probe the combined effect of two stimuli, through extensive ab-initio calculations and Monte-Carlo simulations of the 2D Ising Model, to uncover the magnetic response of CrOCl. First, we establish how the application of uniaxial and biaxial strain can lead to room temperature ferromagnetism and the occurrence of a well-defined phase transition. Next, we show how coupling strain with electric field leads to high temperature magnetic ordering, increasing the T$_c$ to almost three times the intrinsic T$_c$ of CrOCl \cite{pccp4, icee4, aps12}. Further, we have observed spin-selective transport behaviour offering significant spin injection efficiency tunable with applied strain in it \cite{jpcm2}. While extending this work for another member of this family CrOBr, we observed similar tunability in electronic, magnetic phases as well as in the spin-transport behaviour \cite{pe1}. We reported intrinsic magnetism in several new members of the family of TMXYZ system where TM=transition metal and X, Y, Z = Cl, Br, I. It includes robust half-metallicity in two-dimensional VClI$_2$ \cite{jap11}, ferromagnetic semiconductor VClBr$_2$ \cite{jap10}, Janus monolayers \cite{jmr1, aps7}, two-dimensional magnetic semiconductor VIBr$_2$ \cite{cms1} etc. The stimuli assisted tuning and control of magnetic and electronic properties were also studied for Cr$_2$Ge$_2$Se$_6$ \cite{jpd1, icee2, aps6} and Cr$_2$Ge$_2$Te$_6$ \cite{rsc1}.

\section{Twistronics}

The recent discovery of superconductivity in bilayer graphene has fuelled interest in understanding the role of an interlayer twist in controlling the material properties. When two neighbouring layers of a 2D system are rotated with respect to each others, it results in strong-electron correlation effect resulting in the demonstration of novel quantum effects. In this direction, we have studied several interesting systems to understand the role of interlayer twist on the electronic, magnetic and optical properties of various van der Waals heterostructures. Twist-assisted tunability and enhanced ferromagnetism in a 2D Van Der Waal's Heterostructure made up of CrI$_3$ and Graphene \cite{aps3}, while Proximity induced exchange coupling in a Phosphorene heterojunction was reported in the presence of a CrI$_3$ substrate \cite{aps5}. Interesting effects were observed when a 2D magnet interface was formed with a magnetic topological insulator like MnBi$_2$Te$_4$, where topological features and Dirac to Weyl type of band conversion was observed \cite{aps8}. Twist-assisted optoelectronic phase control was reported in two-dimensional (2D) Janus heterostructures that exhibit direct bandgap with type-II band alignment at some specific twist angle, which shows potential for future photovoltaic devices \cite{sr1, aps11}. Similar studies were also performed on different TMD heterostructure like MoS$_2$/WS$_2$ heterostructure \cite{ccm1, aps10}, MoS$_2$/MoSe$_2$ heterostructure \cite{apa4}.

\section{Energy Storage}
The effect of global warming is very prominent and the focus is towards green energy. In this direction, several technologies are in focus like photovoltaics, thermoelectrics, battery storage etc. The use of 2D materials and their heterostructures can have useful advantages due to their larger areal coverage in a smaller volume. We have studied MnO$_2$/CoO$_2$ heterostructure as Promising cathode material  for the Li and Na ion battery \cite{jap12}. Ultralow lattice thermal conductivity and thermoelectric performance were reported for twisted Graphene/Boron Nitride heterostructure through strain engineering \cite{carbon3, aps15} and two-dimensional KCuX (X= S, Se) \cite{aps14}.

\section{Resistive Switching and Non-volatile memory}

In the present era, the demand of Non-volatile memory (NVM) is continuously increasing because of its use in portable gadgets and consumer electronics like, laptops, digital cameras, USB storage devices, and smart phones etc. A variety of NVM candidates have emerged such as Flash memory, Phase change memory (PCM), Ferroelectric random access memory (FeRAM), Magnetic random access memory (MRAM), and Resistive random access memory (RRAM) etc. Among them, RRAM is the most promising candidate for the future memory design due to its simple structure, high endurance, good retention, low operating voltage, high power and multi-functionalities like complementary (CRS), unipolar (URS) and bipolar resistive switching (BRS).

Choice of material in such structures is very crucial which determines the operational limitations in terms of durability, retention, reliability and switching power etc. Till now, a variety of materials like, binary transition metal oxides, organic materials, graphene derivative, chalcogenides, complex compositional materials and polymers have been used for the resistive switching (RS) purpose. Out of these, the transition metal oxide-based devices have attracted significant attention due to their easy fabrication technique and having a wide range of electrical properties. Among metal-oxides, ZnO has an advantage in terms of cost and stability, offering a wide and direct band gap, controllable electrical behaviour, having different morphologies and environmental friendly nature which we have studied in the presence of an oxide electrode \cite{cap1}. Another approach of improving the switching behaviour is through the formation of a hybrid structures made of various 2D materials and perovskite oxides, where we have studied a range of perovskite oxides like LSMO, LBMO, LCMO etc. in the presence of 2D materials like rGO, CuI, Graphene, ZnO etc \cite{jalcom1, jap9, mtc1, mrb1, ml1, apa3, apa2}. To understand the temperature effect on the switching study, studies were performed \cite{apa1, ml2, aip1}. The addition of biological materials as a switching medium also offered useful insight \cite{aps12, aps2}. Recently, we have started looking a supramolecular gel materials for non-volatile memory applications which are found to offer interesting switching properties \cite{aem1, langmuir1}.

\bibliographystyle{unsrt}

\begin{thebibliography}{11}
\section{References}




\bibitem{prl1}
{Heron, David Owen Goudie, Soumya Jyoti Ray, S. J. Lister, C. M. Aegerter, H. Keller, P. H. Kes, G. I. Menon, and Stephen Leslie Lee. ``Muon-spin rotation measurements of an unusual vortex-glass phase in the layered superconductor Bi2.15Sr1.85CaCu2O8+$\delta$." Physical Review Letters 110, no. 10 (2013): 107004}

\bibitem{prb4}
{Khasanov, R., Takeshi Kondo, M. Bendele, Yoichiro Hamaya, A. Kaminski, S. L. Lee, S. J. Ray, and Tsunehiro Takeuchi. ``Suppression of the antinodal coherence of superconducting (Bi, Pb)2(Sr, La)2CuO6+$\delta$ as revealed by muon spin rotation and angle-resolved photoemission." Physical Review B 82, no. 2 (2010): 020511}

\bibitem{prb2}
{Ray, S. J., A. S. Gibbs, S. J. Bending, P. J. Curran, Egor Babaev, C. Baines, A. P. Mackenzie, and S. L. Lee. ``Muon-spin rotation measurements of the vortex state in Sr2RuO4: Type-1.5 superconductivity, vortex clustering, and a crossover from a triangular to a square vortex lattice." Physical Review B 89, no. 9 (2014): 094504}

\bibitem{pss1}
{Ray, Soumya J., and Lambert Alff. ``Superconductivity and Dirac fermions in 112-phase pnictides." physica status solidi (b) 254, no. 1 (2017): 1600163}

\bibitem{prb1}
{Retzlaff, Reiner, Alexander Buckow, Philipp Komissinskiy, Soumya Ray, Stefan Schmidt, Holger Muhlig, Frank Schmidl, Paul Seidel, Jose Kurian, and Lambert Alff. ``Superconductivity and role of pnictogen and Fe substitution in 112-LaPdxPn2 (Pn= Sb, Bi)." Physical Review B 91, no. 10 (2015): 104519}



\bibitem{prb3}
{Flokstra, M. G., S. J. Ray, S. J. Lister, J. Aarts, H. Luetkens, T. Prokscha, A. Suter, E. Morenzoni, and S. L. Lee. ``Measurement of the spatial extent of inverse proximity in a Py/Nb/Py superconducting trilayer using low-energy muon-spin rotation." Physical Review B 89, no. 5 (2014): 054510}

\bibitem{pc1}
{Ray, S. J., S. J. Lister, S. L. Lee, Olav Hellwig, and J. Stahn. ``Stoner enhanced paramagnetic influence on superconductivity in a superconductor/metallic heterostructure." Physica C: Superconductivity 487 (2013): 67-71}


\bibitem{apl1}
{Lister, Stephen J., Tom Thomson, Joachim Kohlbrecher, K. Takano, V. Venkataramana, Soumya Joyti Ray, M. P. Wismayer et al. ``Size-dependent reversal of grains in perpendicular magnetic recording media measured by small-angle polarized neutron scattering." Applied Physics Letters 97, no. 11 (2010): 112503}

\bibitem{jap1}
{Lister, S. J., M. P. Wismayer, V. Venkataramana, M. A. De Vries, S. J. Ray, S. L. Lee, T. Thomson et al. ``Small-angle polarized neutron studies of perpendicular magnetic recording media." Journal of Applied Physics 106, no. 6 (2009): 063908}


\bibitem{cej1}
{Chahal, Sumit, Akhil K. Nair, Soumya Jyoti Ray, Jiabao Yi, Ajayan Vinu, and Prashant Kumar. ``Microwave flash synthesis of phosphorus and sulphur ultradoped graphene." Chemical Engineering Journal 450 (2022): 138447}

\bibitem{pccp6}
{Vishwakarma, Kavita, Shivani Rani, Sumit Chahal, Chia-Yen Lu, Soumya Jyoti Ray, Chan-Shan Yang, and Prashant Kumar. ``Quantum coupled borophene based heterolayers for excitonic and molecular sensing applications." Physical Chemistry Chemical Physics 24, 12816-12826 (2022)}


\bibitem{carbon1}
{Rani, S., and S. J. Ray. ``DNA and RNA detection using graphene and hexagonal boron nitride based nanosensor." Carbon 173 (2021): 493-500.}

\bibitem{icee1}
{Rani, Shivani, and Soumya Jyoti Ray. ``Graphene and hexagonal boron nitride based nano-electronic biosensor." In 2020 5th IEEE International Conference on Emerging Electronics (ICEE), pp. 1-4. IEEE, 2020.}


\bibitem{jap8}
{Mishra, S., S. Rani, and S. J. Ray. ``Single electron transistor based nanosensor for DNA and RNA detection." Journal of Applied Physics 128, no. 19 (2020): 194302.}

\bibitem{sab1}
{Ray, S. J. ``First-principles study of MoS2, phosphorene and graphene based single electron transistor for gas sensing applications." Sensors and Actuators B: Chemical 222 (2016): 492-498.}

\bibitem{jpcm1}
{Ray, S. J., M. Venkata Kamalakar, and R. Chowdhury. ``Ab initio studies of phosphorene island single electron transistor." Journal of Physics: Condensed Matter 28, no. 19 (2016): 195302.}

\bibitem{carbon2}
{Rani, S., and S. J. Ray. ``Detection of gas molecule using C3N island single electron transistor." Carbon 144 (2019): 235-240.}

\bibitem{itn1}
{Gaurav, Kumar, Boddepalli SanthiBhushan, Soumya J. Ray, and Anurag Srivastava. ``Acridinium Based Organic Molecular Single Electron Transistor for High Performance Switching Applications." IEEE Transactions on Nanotechnology 18 (2019): 1148-1155.}

\bibitem{ne1}
{Rani, Shivani, and Soumya Jyoti Ray. ``Detection of CO using a graphene-island-based single-electron transistor." Nanomaterials and Energy 8, no. 2 (2019): 135-138.}

\bibitem{jap7}
{Ray, S. J. ``Gate engineered performance of single molecular transistor." Journal of Applied Physics 119, no. 20 (2016): 204302.}

\bibitem{jap6}
{Ray, S. J. ``Humidity sensor using a single molecular transistor." Journal of Applied Physics 118, no. 4 (2015): 044307.}

\bibitem{jap5}
{Ray, S. J. ``Single molecular transistor as a superior gas sensor." Journal of Applied Physics 118, no. 3 (2015): 034303.}

\bibitem{jap4}
{Ray, S. J. ``Single molecule transistor based nanopore for the detection of nicotine." Journal of Applied Physics 116, no. 24 (2014): 244307.}

\bibitem{jap3}
{Ray, S. J. ``Single atom impurity in a single molecular transistor." Journal of Applied Physics 116, no. 15 (2014): 154302.}

\bibitem{jap2}
{Ray, S. J., and Rajib Chowdhury. ``Double gated single molecular transistor for charge detection." Journal of Applied Physics 116, no. 3 (2014): 034307.}

\bibitem{aip2}
{Rani, Shivani, and S. J. Ray. ``Biosensing using C$_3$N nanoribbon." In AIP Conference Proceedings, vol. 2265, no. 1, p. 030706. AIP Publishing LLC, 2020.}

\bibitem{pccp1}
{Rani, S., and S. J. Ray. ``Two-dimensional C$_3$N based sub-10 nanometer biosensor." Physical Chemistry Chemical Physics 22, no. 20 (2020): 11452-11459.}

\bibitem{pra1}
{Chaudhury, A., S. Majumder, and S. J. Ray. ``Proximity-induced colossal conductivity modulation in phosphorene." Physical Review Applied 11, no. 2 (2019): 024056.}



\bibitem{pra2}
{Clapera, Paul, Soumya Ray, Xavier Jehl, M. Sanquer, A. Valentian, and Sylvain Barraud. ``Design and cryogenic operation of a hybrid quantum-CMOS circuit." Physical Review Applied 4, no. 4 (2015): 044009}

\bibitem{cpem1}
{Charron, T., L. Devoille, S. Djordjevic, O. Soron, F. Piquemal, P. Clapera, S. J. Ray, X. Jehl, R. Wacquez, and M. Vinet. ``Characterization of hybrid metal/semiconductor electron pumps for quantum metrology." In 29th Conference on Precision Electromagnetic Measurements (CPEM 2014), pp. 442-443. IEEE, 2014}

\bibitem{wolte1}
{Clapera, Paul, Soumya Ray, Xavier Jehl, Marc Sanquer, Alexandre Valentian, and Sylvain Barraud. ``A quantum device driven by an on-chip CMOS ring oscillator." In 2014 11th International Workshop on Low Temperature Electronics (WOLTE), pp. 73-76. IEEE, 2014}

\bibitem{cpem2}
{Ray, S. J., P. Clapera, X. Jehl, T. Charron, S. Djordjevic, L. Devoille, E. Potanina, G. Barinovs, and V. Kashcheyevs. ``Modeling of an adiabatic tunable-barrier electron pump." In 29th Conference on Precision Electromagnetic Measurements (CPEM 2014), pp. 446-447. IEEE, 2014}

\bibitem{prx1}
{Jehl, X., B. Voisin, T. Charron, P. Clapera, S. Ray, B. Roche, M. Sanquer et al. ``Hybrid metal-semiconductor electron pump for quantum metrology." Physical Review X 3, no. 2 (2013): 021012}




\bibitem{apr1}
{Dayen, Jean-Francois, Soumya J. Ray, Olof Karis, Ivan J. Vera-Marun, and M. Venkata Kamalakar. ``Two-dimensional van der Waals spinterfaces and magnetic-interfaces." Applied Physics Reviews 7, no. 1 (2020): 011303}

\bibitem{aps4}
Kumari, Puja, and Soumya Jyoti Ray. ``Tunable Optical Properties in Magnetic Phosphorene." In APS March Meeting Abstracts, vol. 2022, pp. T00-220. 2022.

\bibitem{mtp1}
{Kumari, Puja, and Soumya Jyoti Ray. ``The magnetic anisotropy and spin filtering effect in ferromagnetic phosphorene." Materials Today: Proceedings 67 (2022): 931-934}

\bibitem{aip3}
{Nair, Akhil K., Puja Kumari, and Soumya Jyoti Ray. ``High temperature magnetic ordering in manganese doped phosphorene nanoribbon." In AIP Conference Proceedings, vol. 2265, no. 1, p. 030688. AIP Publishing LLC, 2020}

\bibitem{jpcc1}
{Chahal, Sumit, Tumesh Kumar Sahu, Subhasmita Kar, Soumya J. Ray, Vasudevanpillai Biju, and Prashant Kumar. ``Transition Metal-Doped Boron Nitride Atomic Sheets with an Engineered Bandgap and Magnetization." The Journal of Physical Chemistry C 126, no. 49 (2022): 21084-21093}

\bibitem{pccp3}
{Kumari, P., S. Majumder, S. Rani, A. K. Nair, K. Kumari, M. Venkata Kamalakar, and S. J. Ray. ``High efficiency spin filtering in magnetic phosphorene." Physical Chemistry Chemical Physics 22, no. 10 (2020): 5893-5901}

\bibitem{pccp2}
{Nair, A. K., P. Kumari, M. Venkata Kamalakar, and S. J. Ray. ``Dramatic magnetic phase designing in phosphorene." Physical Chemistry Chemical Physics 21, no. 42 (2019): 23713-23719}

\bibitem{aps13}
Kumari, Puja, and Soumya Jyoti Ray. ``Tunable Optical Properties in Magnetic Phosphorene." Bulletin of the American Physical Society (2022).

\bibitem{pccp5}
{Ray, S. J., and M. Venkata Kamalakar. ``Unconventional strain-dependent conductance oscillations in pristine phosphorene." Physical Chemistry Chemical Physics 20, no. 19 (2018): 13508-13516}

\bibitem{pccp4}
{Nair, A. K., Shivani Rani, M. Venkata Kamalakar, and Soumya Jyoti Ray. ``Bi-stimuli assisted engineering and control of magnetic phase in monolayer CrOCl." Physical Chemistry Chemical Physics 22, no. 22 (2020): 12806-12813}


\bibitem{icee4}
{Kar, Subhasmita, Shivani Rani, Akhil Nair, M. Venkata Kamalakar, and Soumya Jyoti Ray. ``Tuning and control of the magnetic properties of a 2D nanomagnet." In 2020 5th IEEE International Conference on Emerging Electronics (ICEE), pp. 1-4. IEEE, 2020}

\bibitem{aps12}
Rahaman, Towhidur, SHANTANU MAJUMDER, and Soumya Ray. ``Microwave-synthesis of Two-Dimensional Metal Oxychloride Using Metal Chloride Precursor: Experimental and Theoretical Study." Bulletin of the American Physical Society (2023).

\bibitem{jpcm2}
{Rani, Shivani, A. K. Nair, M. Venkata Kamalakar, and Soumya Jyoti Ray. ``Spin-selective response tunability in two-dimensional nanomagnet." Journal of Physics: Condensed Matter 32, no. 41 (2020): 415301}

\bibitem{pe1}
{Kar, S., S. Rani, and S. J. Ray. ``Stimuli assisted electronic, magnetic and optical phase control in CrOBr monolayer." Physica E: Low-dimensional Systems and Nanostructures 143 (2022): 115332}

\bibitem{jap11}
{Mukherjee, T., P. Kumari, S. Kar, C. Datta, and S. J. Ray. ``Robust half-metallicity and tunable ferromagnetism in two-dimensional VClI2." Journal of Applied Physics 133, no. 8 (2023): 084303.}

\bibitem{jap10}
{Kumari, Puja, Tania Mukherjee, Subhasmita Kar, and S. J. Ray. ``VClBr2: A new two-dimensional (2D) ferromagnetic semiconductor." Journal of Applied Physics 133, no. 18, (2023): 183901}

\bibitem{jmr1}
{Mukherjee, T., Kar, S. and Ray, S.J., 2022. Two-dimensional Janus monolayers with tunable electronic and magnetic properties. Journal of Materials Research, 37, 3418–3427 (2022)}

\bibitem{aps7}
Kumar, Saurav, Subhasmita Kar, and Soumya Ray. ``Stimuli controlled electronic and magnetic properties of two-dimensional (2D) magnetic Janus materials." Bulletin of the American Physical Society (2023).

\bibitem{cms1}
{Mukherjee, T., S. Kar, and S. J. Ray. ``Tunable electronic and magnetic properties of two-dimensional magnetic semiconductor VIBr2." Computational Materials Science 209 (2022): 111319}

\bibitem{jpd1}
{Kar, S., A. K. Nair, and S. J. Ray. ``Supreme enhancement of ferromagnetism in a spontaneous-symmetry-broken 2D nanomagnet." Journal of Physics D: Applied Physics 54, no. 10 (2020): 105001}

\bibitem{icee2}
{Kar, Subhasmita, Akhil Nair, and Soumya Jyoti Ray. ``Electronic and magnetic phase transition in monolayer Cr2Ge2Se6." In 2020 5th IEEE International Conference on Emerging Electronics (ICEE), pp. 1-4. IEEE, 2020.}

\bibitem{aps6}
Kar, Subhasmita, Akhil Nair, and Soumya Ray. ``Stimuli-assisted magnetism in two-dimensional (2D) magnets." Bulletin of the American Physical Society (2023).

\bibitem{rsc1}
{Nair, A. K., and S. J. Ray. ``Electronic phase-crossover and room temperature ferromagnetism in a two-dimensional (2D) spin lattice." RSC advances 11, no. 2 (2021): 946-952}

\bibitem{aps3}
Subhasmita Kar, and Soumya Jyoti Ray. ``Twist-assisted tunability and enhanced ferromagnetism in a 2D Van Der Waal's Heterostructure." In APS March Meeting Abstracts, vol. 2022, pp. T00-218. 2022.

\bibitem{aps5}
Kumari, Puja, and Soumya Ray. ``Proximity induced exchange coupling in a Phosphorene heterojunction with CrI 3 substrate: a first-principles calculation." Bulletin of the American Physical Society (2023).

\bibitem{aps8}
MAJUMDER, SHANTANU, and Soumya Ray. ``Magnetic and transport study of magnetic topological insulators." Bulletin of the American Physical Society (2023).

\bibitem{sr1}
{Kar, S., P. Kumari, M. Venkata Kamalakar, and S. J. Ray. ``Twist-assisted optoelectronic phase control in two-dimensional (2D) Janus heterostructures." Scientific Reports 13, no. 1 (2023): 13696.}

\bibitem{aps11}
Kar, Subhasmita, Puja Kumari, and Soumya Ray. ``Twist Induced Tunability of Electronic and Optical Properties in a Van der Waal Janus Heterostructure." Bulletin of the American Physical Society (2023).


\bibitem{ccm1}
{Sachin, Saurav, Puja Kumari, Neelam Gupta, Shivani Rani, Subhasmita Kar, and Soumya Jyoti Ray. ``Van der Waals twistronics in a MoS2/WS2 heterostructure." Computational Condensed Matter 35 (2023): e00797}

\bibitem{aps10}
Sahoo, Shubham, Saurav Sachin, Shivani Rani, Puja Kumari, Subhasmita Kar, and Soumya Ray. ``Van der Waals twistronics in a MoS2/WS2 heterostructure." Bulletin of the American Physical Society (2023).

\bibitem{ccm1}
{Sachin, S., Kumari, P., Gupta, N., Rani, S., Kar, S., \& Ray, S. J. (2023). Van der Waals twistronics in a MoS2/WS2 heterostructure. Computational Condensed Matter, 35, e00797}

\bibitem{apa4}
{Sachin, S., Rani, S., Kumari, P., Kar, S., Ray, S. J., Twist-engineered tunability in vertical MoS2/MoSe2 heterostructure. Applied Physics A, 129(1), 46 (2023)}


\bibitem{jap12}
{Sahoo, Shubham, Puja Kumari, and Soumya Jyoti Ray. ``Promising cathode material MnO2/CoO2 heterostructure for the Li and Na ion battery: A computational study." Journal of Applied Physics 134, no. 10 (2023): 104302}.

\bibitem{carbon3}
{Gupta, Neelam, Shivani Rani, Puja Kumari, Rajeev Ahuja, and Soumya Jyoti Ray. "Ultralow lattice thermal conductivity and thermoelectric performance of twisted Graphene/Boron Nitride heterostructure through strain engineering." Carbon (2023): 118437}

\bibitem{aps15}
Rani, Shivani, and Soumya Ray. ``Ultralow lattice thermal conductivity of Gr/h-BN heterostructure using controlled twist engineering." Bulletin of the American Physical Society (2023).

\bibitem{aps14}
Gupta, Neelam, Shubham Kumar, Shivani Rani, Soumya Ray, and Puja Kumari. ``Ultralow thermal conductivity and thermoelectric performance of two-dimensional KCuX (X= S, Se) materials for energy harvesting." Bulletin of the American Physical Society (2023).


\bibitem{cap1}
{Kumari, Karuna, Subhasmita Kar, Ajay D. Thakur, and S. J. Ray. ``Role of an oxide interface in a resistive switch." Current Applied Physics 35 (2022): 16-23.}

\bibitem{ml1}
{Majumder, Shantanu, Karuna Kumari, and S. J. Ray. ``Pulsed voltage induced resistive switching behavior of copper iodide and La0. 7Sr0. 3MnO3 nanocomposites." Materials Letters 302 (2021): 130339.}

\bibitem{mrb1}
{Kumari, Karuna, Ashutosh Kumar, Ajay D. Thakur, and S. J. Ray. ``Charge transport and resistive switching in a 2D hybrid interface." Materials Research Bulletin 139 (2021): 111195.}

\bibitem{mtc1}
{Kumari, Karuna, Ajay D. Thakur, and S. J. Ray. ``The effect of graphene and reduced graphene oxide on the resistive switching behavior of La0. 7Ba0. 3MnO3." Materials Today Communications 26 (2021): 102040.}

\bibitem{jap9}
{Kumar, Ashutosh, Karuna Kumari, S. J. Ray, and Ajay D. Thakur. ``Graphene mediated resistive switching and thermoelectric behavior in lanthanum cobaltate." Journal of Applied Physics 127, no. 23 (2020): 235103.}

\bibitem{apa3}
{Kumari, Karuna, S. J. Ray, and Ajay D. Thakur. ``Resistive switching phenomena: a probe for the tracing of secondary phase in manganite." Applied Physics A 128, no. 5 (2022): 1-11.}

\bibitem{apa2}
{Kumari, Karuna, Ajay D. Thakur, and S. J. Ray. ``Structural, resistive switching and charge transport behaviour of (1-x)La0.7Sr0.3MnO3.(x)ZnO composite system." Applied Physics A 128, no. 11 (2022): 992.}

\bibitem{jalcom1}
{Kumari, Karuna, Ashutosh Kumar, Dinesh K. Kotnees, Jayakumar Balakrishnan, Ajay D. Thakur, and S. J. Ray. ``Structural and resistive switching behaviour in lanthanum strontium manganite-reduced graphene oxide nanocomposite system." Journal of Alloys and Compounds 815 (2020): 152213.}


\bibitem{ml2}
{Kumari, Karuna, S. Majumder, Ajay D. Thakur, and S. J. Ray. ``Temperature-dependent resistive switching behaviour of an oxide memristor." Materials Letters 303 (2021): 130451.}


\bibitem{aip1}
{Kumari, Karuna, Ashutosh Kumar, Ajay D. Thakur, and S. J. Ray. ``Effect of temperature and magnetic field in resistive switching behavior of La0. 7Ca0. 3MnO3. rGO nano-composite." In AIP Conference Proceedings, vol. 2220, no. 1, p. 080007. AIP Publishing LLC, 2020.}

\bibitem{apa1}
{Majumder, S., K. Kumari, and S. J. Ray. ``Temperature-dependent resistive switching behavior of a hybrid semiconductor-oxide planar system." Applied Physics A 129, no. 5 (2023): 1-10.}


\bibitem{aps12}
Kumari, Dimpal, and Soumya Ray. ``Resistive switching in a Bio-Resistive Random-Access Memory device (Bio-ReRAM)." Bulletin of the American Physical Society (2023).

\bibitem{aps2}
Roy, Arpita, and Soumya Ray. ``Temperature dependent resistive switching behavior of a pectin based heterostructure." Bulletin of the American Physical Society (2023).

\bibitem{aem1}
{Karmakar, Kripasindhu, Arpita Roy, Subhendu Dhibar, Shantanu Majumder, Subham Bhattacharjee, Bijnaneswar Mondal, SK Mehebub Rahaman, Ratnakar Saha, Soumya Jyoti Ray, and Bidyut Saha. ``Instantaneous Gelation of a Self-Healable Wide-Bandgap Semiconducting Supramolecular Mg (II)-Metallohydrogel: An Efficient Nonvolatile Memory Design with Supreme Endurance." ACS Applied Electronic Materials 2023, 5(6), 3340–3349}




\bibitem{langmuir1}
{Alam, Noohul, Shantanu Majumder, Soumya Jyoti Ray, and Debajit Sarma. ``A Wide Bandgap Semiconducting Magnesium Hydrogel: Moisture Harvest, Iodine Sequestration, and Resistive Switching." Langmuir 38, no. 34 (2022): 10601-10610.}









\end{thebibliography}

\end{document}